\documentstyle[11pt,paspconf,epsfig]{article}

\newbox\grsign \setbox\grsign=\hbox{$>$} \newdimen\grdimen \grdimen=\ht\grsign
\newbox\simlessbox \newbox\simgreatbox \newbox\simpropbox
\setbox\simgreatbox=\hbox{\raise.5ex\hbox{$>$}\llap
     {\lower.5ex\hbox{$\sim$}}}\ht1=\grdimen\dp1=0pt
\setbox\simlessbox=\hbox{\raise.5ex\hbox{$<$}\llap
     {\lower.5ex\hbox{$\sim$}}}\ht2=\grdimen\dp2=0pt
\setbox\simpropbox=\hbox{\raise.5ex\hbox{$\propto$}\llap
     {\lower.5ex\hbox{$\sim$}}}\ht2=\grdimen\dp2=0pt
\def\gtrsim{\mathrel{\copy\simgreatbox}}
\def\lesssim{\mathrel{\copy\simlessbox}}

\begin{document}

\title{Relativistic Outflows in Gamma Ray Bursts}

\author{Boris Stern}
\affil{Institute for Nuclear Research, Moscow 117312, Russia}
\affil{and}
\affil{Stockholm Observatory, S-133 36 Saltsj\"obaden, Sweden}

\begin{abstract}
Despite that gamma-ray bursts is a phenomenon quite different from accreting 
compact objects, it could be that their hard x-ray emission is 
associated with a very similar
mechanism of energy dissipation. In both cases, we could deal with 
reconnection of a turbulent magnetic field with intensive $e^+e^-$ pair 
production and quasi-thermal Comptonization.
\end{abstract}


\section{Introduction}

 The approximate consensus on gamma ray bursts (GRBs) can be reduced 
to a few brief statements:

-- GRBs are cataclysmic events with an energy release $\sim 10^{51}$erg in 
$\gamma$-rays (assuming isotropic emission) at cosmological distances.

-- The primary event is a coalescence of two compact objects of stellar origin
(neutron stars and black holes, Blinnikov et al. 1984; Paczy\'nski 1992) 
or an exotic explosion of a single 
stellar object (hypernova, Paczy\'nski  1998)

-- All we see are effects associated with an expanding blast wave (fireball),
or propagating jet,
or multiple colliding shocks of dimensions and time scales of a few order  
of magnitude larger than the scales of the primary event (which is invisible in 
itself at the present level of sensitivity, see M\'esz\'aros \& Rees 1993).

There exist other points of view, of course, e.g., GRBs as sporadic 
microblazars (Shaviv \& Dar 1995a; for criticism of fireball models, see 
Dar 1998). 
I will use the fireball paradigm,
keeping in mind a jet geometry as an alternative. The principal problems arising 
in the inhomogeneous fireball and the jet scenarios as well as the possible 
underlying physical processes are similar.

  There are two classes of GRBs (which could be different phenomena or 
different modes of the same phenomenon) -- 
short ($\lesssim 1$ s ) and long ($\gtrsim$ 1 s). The discussion below 
concerns the long GRBs. 
 Due to the large intensity of many bursts we have very rich hard X-ray/soft 
$\gamma$-ray GRB data: excellent light curves and good spectra. However the 
data are so diverse and sometimes puzzling, that usually new good data
complicate the problem rather than clarify it.  
I will start with the time variability data and try to review possible
conclusions inferred from temporal properties. Then I will review the spectral
properties and discuss possible regimes of emission.

\section{Time Variability, Phenomenology}

 GRB temporal properties which are worth to emphasize are the following:

 2.1. {\bf Bimodality}. The duration distribution of GRBs extending over 5 
orders of magnitude has two humps (Meegan et al. 1998) which are 
believed to correspond to different classes of GRBs: short and long GRBs,
separated by a minimum around 2 s. These could also  just be different modes 
of the same phenomenon (sometimes a precursor looking like a short burst is 
followed by a long burst, Fig.~1e). 
Our discussion concerns mainly long bursts which
constitute 70 - 75 \% of GRBs

\begin{figure}
\centerline{\epsfig{file=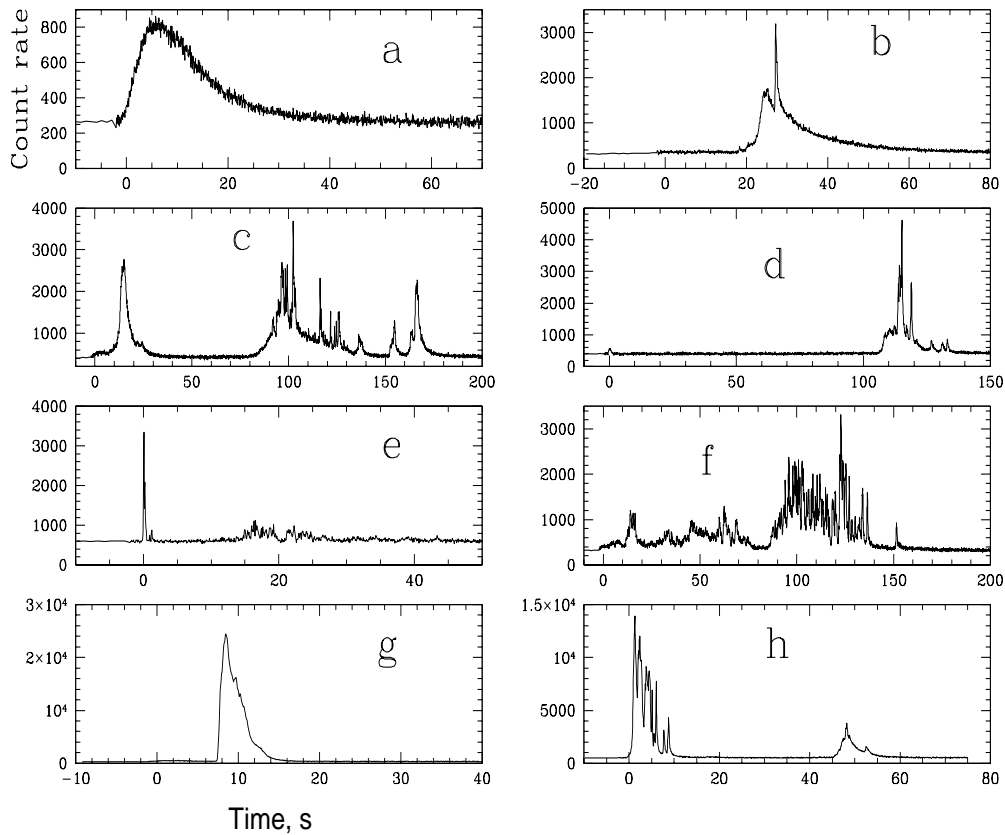,width=14cm,height=12cm}}
\caption{Examples of light curves of strong gamma-ray bursts: a) a GRB consisting of a 
single canonical pulse; b) two overlapping pulses of different durations;
c) an event showing a wide range of time scales; d) an event with 
a weak short precursor; e) a GRB looking like combination of a typical short
and a typical long GRB, there are several events of this kind in the BATSE 
sample; f) one of most complex events; g) the strongest burst of the ``long''
class; h) two episodes separated by a very quiet interval. 
} \label{fig-1}
\end{figure}

 2.2. {\bf Diversity}. Some events consist of a single smooth pulse of almost standard
shape (Fig.~1a), others are very complex and chaotic (Fig.~1f), and
some are a combination of smooth 
pulses and chaotic intervals (Fig.~1c). 
No distinct morphological classes are found.
At first sight, GRB light curves obey no rules.

 2.3. {\bf Large amplitude of variations}. There are strong events with emission 
episodes separated by quiet intervals. The upper limit for emission
between episodes is below $10^{-3}$  of the peak flux in some events (see Figs.~1h, 2h). 
In other terms, the emission can turn off to a very low level and then turn 
on again.

\begin{figure}
\centerline{\epsfig{file=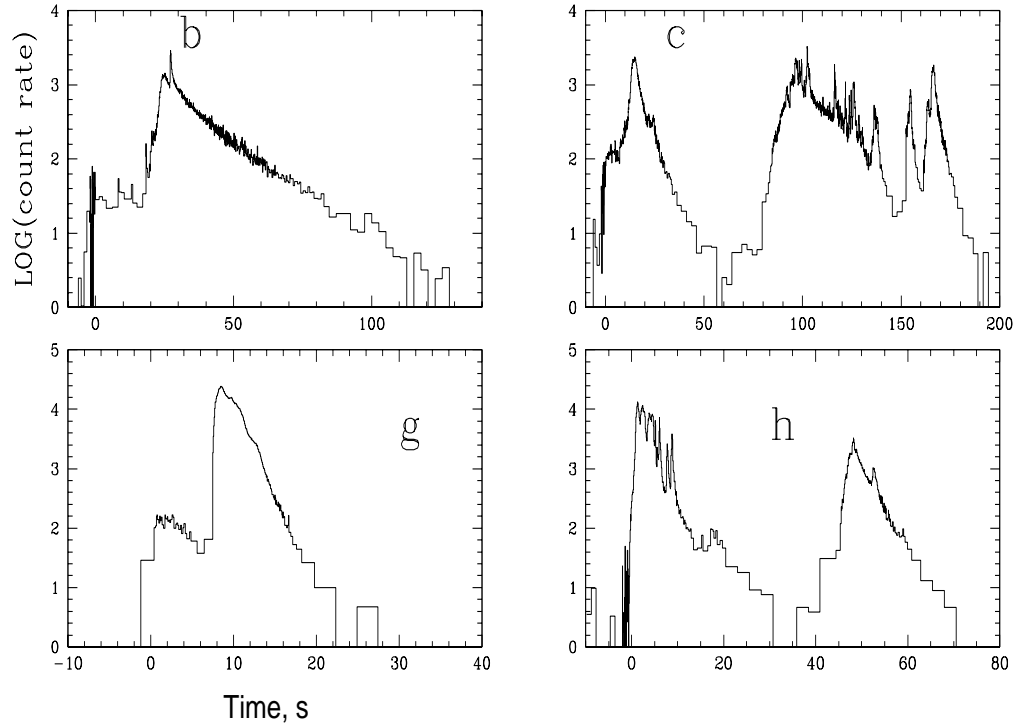,width=14cm,height=10cm}}
\caption{Some of the examples from Fig.~1 with 
subtracted background plotted in logarithmic scale.
Labels correspond to those in Fig.~1. Note the large dynamical range of the 
light curves: near-exponential tails can be traced over almost 4 orders of 
magnitude of intensity (g), the intensity can drop below $10^{-3}$ of the peak 
value and then regain the same level (c, h). Event (g) consists of two peaks
which are morphologically similar and differ in amplitude by a factor of 200. This 
tells us something about intrinsic luminosity function: if the emission
of pulses is a stochastic process, then the weak pulse in event (g) could be 
emitted without the strong one and the event would be 200 times weaker
(but still detectable). 
 Tails of pulses are almost never perfect exponentials. Nevertheless they
are better described as exponentials than as power laws. 
} \label{fig-2}
\end{figure}

 2.4. {\bf  Composite structure}. Any event is the sum of elementary pulses which are additive and can 
overlap.
 This statement is difficult to
prove. It is, however, a stable impression. This is more or less obvious  
for events consisting of a few pulses (Fig.~1b) and seems be a 
reasonable generalization for
chaotic events. The most erratic events could consist of $\sim 1000$ pulses
(Stern \& Svensson 1996). For attempts to decompose bursts into single pulses,
see Norris et al. (1996)

 2.5. {\bf Absence of a starting mark}. There is no typical feature in the 
light curves that could be associated with 
the primary event. A burst can begin in very different ways - a slow smooth 
rise (Fig.~1a,b,f), a sharp abrupt rise (Fig.~1h), a weak precursor 
separated by tens of seconds from the main event (Fig.~1d), etc. 

 2.6. {\bf Weak and slow time evolution}. The direct time evolution of complex 
bursts from their beginning to the end is slow and weak (Fig.~1c,f). 
There are only 
slow statistical trends: hard-to-soft evolution is more frequent than 
soft-to-hard (Ford et al. 1995) and the highest peak of the burst has a 
statistical tendency to appear 
at the beginning. A direct time dependence should exist, but it is not easy 
to extract and its typical characteristic scale exceeds 100 s.

 Summarizing 2.5 and 2.6 we can state that the primary event leaves no 
mark and we cannot define the ``zero time'' for the event. 

 These are phenomenological facts which one can derive from just 
looking at many
time profiles or plotting the simplest distributions. In the next section, 
I will 
consider a more quantitative description of the time variability.

\section{Time Variability: Wide Range of Time-scales and Self-Similarity}

 Stern (1996) found that the average peak aligned profile of all BATSE GRBs
has a stretched exponential (SE) shape:
\begin{equation}
I = I_0 \exp\left[ -(t/t_0)^{1/3}\right] ,
\end{equation}
where $t$ is time since the highest peak of the event, and $t_0$ is a time 
constant $\sim$ 0.5 s. 
This dependence extends over 3 orders of magnitude in time ($0.2 - 200$ s)
and over 2.5 orders in the amplitude of the average signal (Fig.~3). 
It is worth to note
that the similar distribution for solar flares is not such a good SE. 
If one tries to describe the average time profile of solar flares 
with a SE one  obtains
an index close to 1/2 instead of 1/3 (Stern 1996).
Stretched exponentials are quite common in complex dynamical systems with a 
wide spectrum of variations. An example which will be discussed below 
is turbulence where some distributions have an SE 
shape (Ching 1991;  Jensen, Paladin, \& Vulpiani  1992).
 We can speculate that SEs are  associated 
with near-critical systems where the criticality is not complete. In the case of 
exact criticality, the characteristic scale (e.g. the time constant) disappear 
and 
all distributions should convert into a power law. The SE
does contain the time constant $t_0$. It does not mean that we have found 
a characteristic time scale. An SE can be associated with a 
truncated 
power law spectrum. Then $t_0$ is some function of $t_{\min}$ and $t_{\max}$ 
at which the system changes its behavior. 

 Indeed, the average power density spectrum (PDS) of long bursts is a 
truncated
power law, $P(f) \propto f^{-1.67}$ (Beloborodov, Stern \& Svensson 1999),  
extending between 0.02 and 2 Hz (Fig.~4).
The amusing fact is that the average PDS has exactly the same slope $(-5/3)$ 
as the Kolmogorov spectrum describing the energy distribution in developed 
turbulence. This will be discussed below.

\begin{figure}
\centerline{\epsfig{file=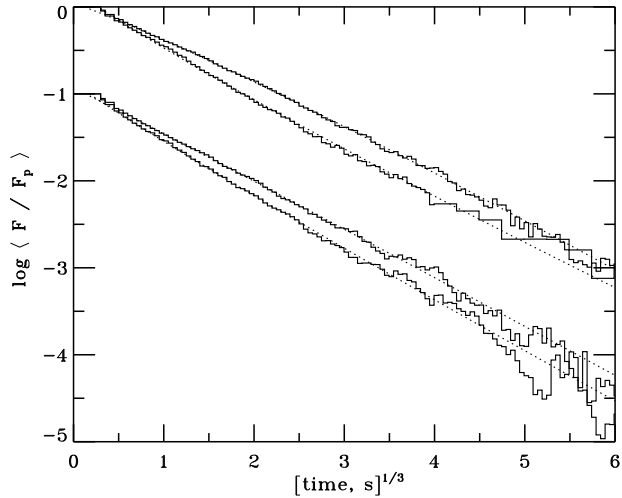,width=9cm}}
\caption{Stretched exponential slopes (rising and decaying) of the 
average time profile 
of GRBs (from Stern, Poutanen \& Svensson 1999). Upper set of curves: the full
useful BATSE 4 sample, 1310 GRBs. Lower set of curves (shifted down): the 953
brightest GRBs. Rising slopes are steeper. Dotted curves represent stretched 
exponential fits.  
} \label{fig-3}
\end{figure}

\begin{figure}
\centerline{\epsfig{file=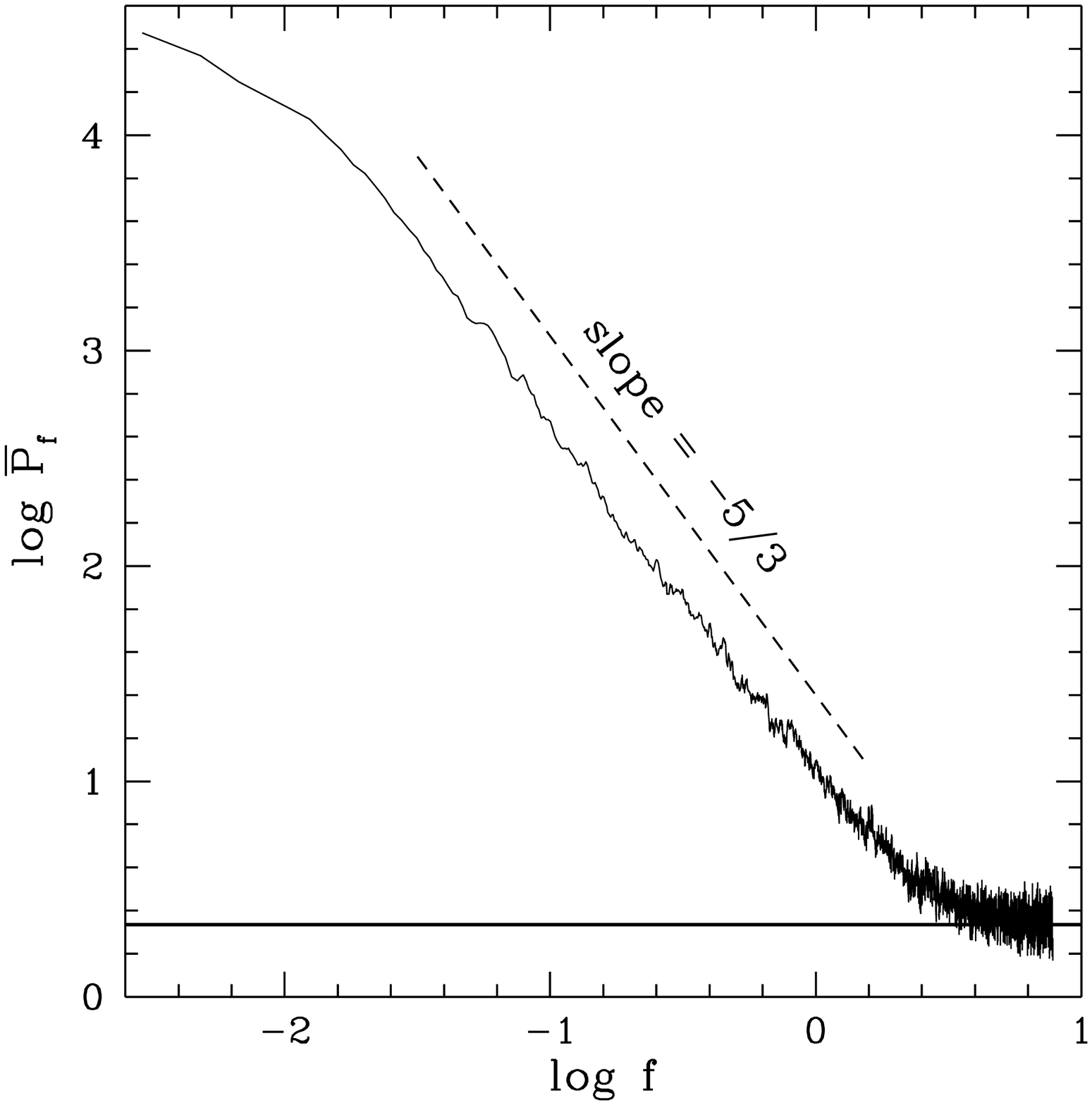,width=8.6cm,height=7.5cm}} 
\centerline{\epsfig{file=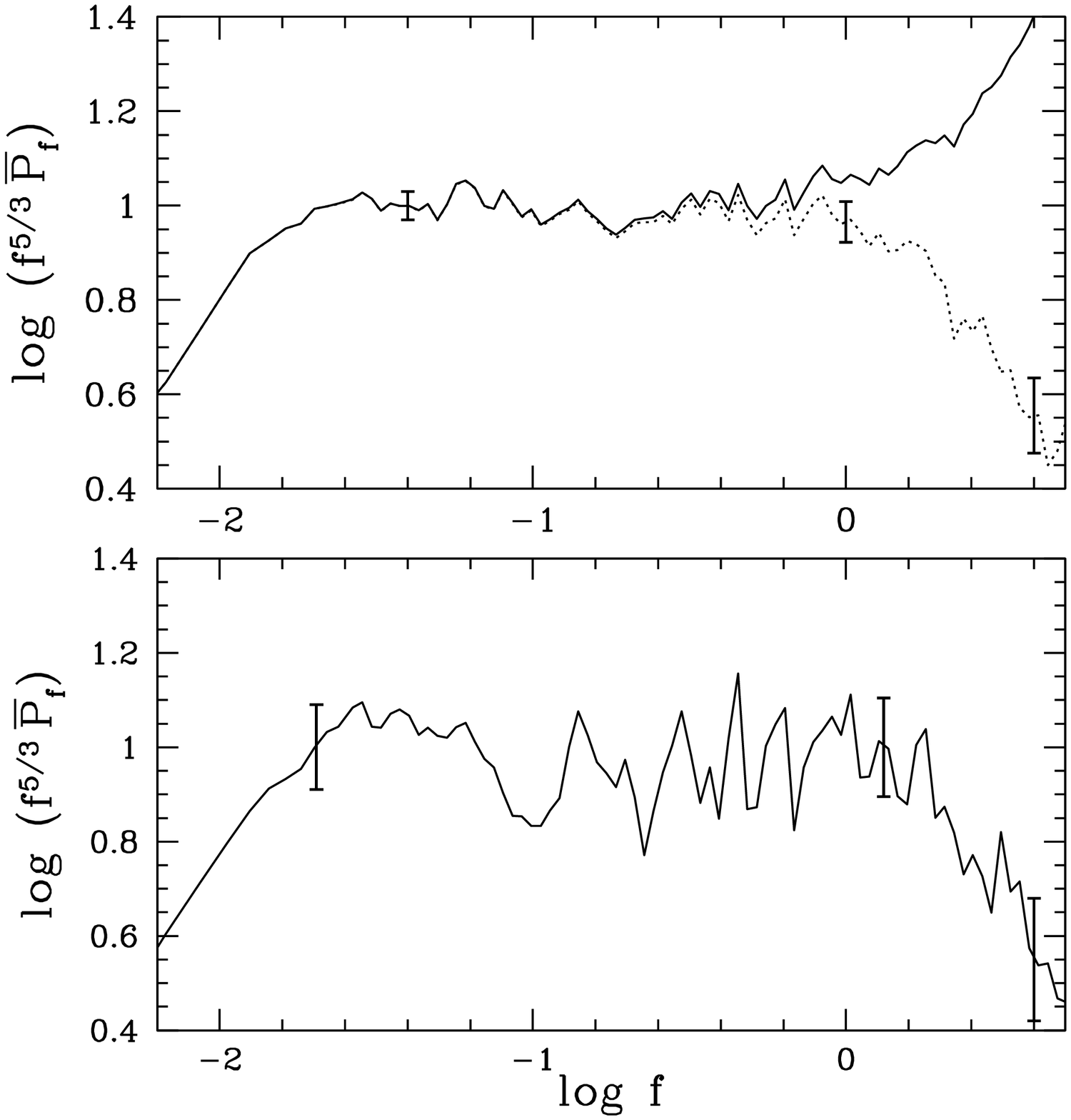,width=9cm}} 
\caption{Average Fourier power density spectrum for 214 long ($T_{90} > 20$ s)
strong bursts (from Beloborodov, Stern \& Svensson 1999). 
Upper panel: the original PDS, the solid horizontal line shows the average
Poisson level. Middle panel: the PDS multiplied by $f^{5/3}$. Dotted curve 
shows the spectrum after subtraction of the Poisson level. Bottom panel:
the average PDS for the 27 strongest events in the sample.
} \label{fig-4}
\end{figure}

The low frequency
turnover is associated with the well known turnover in the duration 
distribution of 
GRBs ($\sim$ 30 s) which in turn is associated
with some global properties of the phenomenon.  The high frequency turnover is 
something new. It should be associated with some nonlinearity in  
the physical processes appearing at a certain scale of the emitting systems.
Maybe this is related to the compactness parameter of local events associated
with the emission of separate pulses. 

 What can we conclude from all these facts? 

\begin{itemize}
\item 
We probably deal with a complex dynamical system which generates
a wide spectrum of features exhibiting some scaling invariance 
(self-similarity) over at least  2 orders of magnitude.  

\item 
Regular extended distributions indicate that all bursts, despite their
diversity,  can be considered 
as different random realizations of the same stochastic process.

\item 
The underlying stochastic process is close to a near-critical regime. 
This is what we need in order to observe a huge diversity of GRBs. 
Otherwise we would have to assume 
very different conditions in different bursts. Near-criticality provides  
large fluctuations under stable conditions.
\end{itemize}

 These conclusions are speculative, of course, and can hardly  be formulated
quantitatively. Nevertheless, a toy pulse avalanche model of Stern \& Svensson
(1996) constructed on the basis of these assumptions gives a successful 
quantitative statistical description of GRBs (including the stretched 
exponential average profile and the power-law average PDS). 
The model in a near critical regime reproduces the diversity of GRBs for the 
same set of parameters.
The success of the 
model does not mean that the pulse avalanche model is valid and is the only 
possibility. It rather means that the approach based the above conclusions 
is reasonable.

\section{Underlying Scenario: A Recurrent  Central Engine 
versus a Turbulent  or Inhomogeneous Fireball}

 What physics can be behind the stochastic process discussed in the 
previous section?

 The best studied scenario of GRB emission is based on a relativistic 
expanding fireball (Cavallo \& Rees 1978) energized by the merging of two
compact objects. If the baryon loading of the fireball is small then it 
must be ultrarelativistic (Pacy\'nski 1990). In an early stage, the 
fireball cannot emit efficiently just because the radiation is trapped 
due to the  very 
large optical depth and almost all energy goes into kinetic
energy (see M\'esz\'aros \& Rees 1993). 
It agrees with fact 2.5 - we do not see any marker of the 
primary event. And later, when the fireball becomes optically thin and 
interacts with the interstellar medium, it emits a GRB through shock 
particle acceleration.   

 This scenario satisfies the energy requirements and can reproduce a proper
time scale of tens of seconds at Lorentz factors $\sim 100 - 300$.
What is missing in this scenario in its straightforward version is
the complex stochastic time behavior with the properties summarized above.

Fenimore, Madras, \& Nayakshin (1996) and Fenimore, Ramires, \& Summers (1998) 
found an evident controversy in the simplest 
model of the single expanding shell. If the expanding relativistic homogeneous 
shell emits an instantaneous flash, the observer will see an extended pulse
with a characteristic width $\sim t-t_0$, where $t$ is the observation time
of the beginning of the pulse and $t_0$ is the observation time of the primary
event (if it were observed) producing the expanding shell. Therefore, if we 
associate a pulse in a GRB with a flash of a single relativistic shell,   
we should see nothing earlier than $t - \Delta t$ where $\Delta t$ is
the characteristic time scale of the pulse. This is apparently not the case
in many complex GRBs.

 Another argument against a single explosion is the low filling factor
(i.e., the ratio of the area of emitting regions to the total fireball surface) 
derived
from the time variability (Fenimore et al. 1998). The low filling factor 
leads  to a low efficiency (Piran \& Sari 1997). 

 These problems gave support to ``recurrent central engine'' models
which have  become very popular (e.g., Rees \& M\'es\'zaros 1994; 
Sari \& Piran 1997). The recurrent central engine is usually
described as a long-living (up to hundreds of seconds) accreting system where
an accretion disc is formed by a disrupted neutron star. The system 
emits relativistic shocks that collide producing pulses of gamma ray emission
(Kobayashi, Piran, \& Sari 1997; Daigne \& Mochkovich 1998).
 
How can we then reproduce a wide power-law PDS from the central  engine?
Probably there is no way to do this with a straightforward internal shock
model. Light curves simulated with internal shocks  have nothing common
with real bursts. They have an intrinsic time constant and a very different
Fourier PDS with a power law asymptotic   with the wrong slope: 
$P(f)={\rm const}$ instead of $P(f)=f^{-5/3}$. 

In principle, an accreting system can provide a power law PDS,
e.g., the Cyg X-1 PDS is a power law $P(f)=f^{-1}$ over 1.5 decades
(from 0.03 Hz to 1 Hz, see Belloni \& Hasinger 1990). 
However, we cannot see the time profile
produced by the central engine itself as the history of accretion will be
reprocessed by 
internal shocks. The Kolmogorov PDS can hardly be obtained straightforwardly 
with internal shocks because too much power has to be 
transferred to low frequencies. Maybe one can invent some rule 
for the ejection of internal shocks to reproduce the Kolmogorov slope.
However, this would be something farfetched.

The long-living central engine helps to solve some problems such as 
a very slow (if existing) evolution of temporal and spectral properties 
in complex bursts. Nevertheless,
we need something else, more complicated than shock collisions, to
produce the self-similar behavior over 2 decades of time scales. 
As was emphasized
in the previous section, we need a complex dynamical process for this.
I would suggest that we should search for such a process in the shock 
evolution rather than in collisions of internal shocks. 

 We can suggest at least two suitable dynamical processes: MHD turbulence,
which is very natural in an relativistic outflow and dynamical instabilities,
most probably the Rayleigh - Taylor instability. Both can generate a wide
range of irregularities with a high energy density contrast. Reconnection of 
the magnetic field generated by a turbulent dynamo is certainly a very  
efficient way to dissipate the energy into gamma rays.

Some arguments in favor of this scenario can be borrowed from solar flares. 
Their time behavior resembles
GRBs (while it still differs from GRBs at a quantitative
level -- solar flares have a different average time profile and a different 
average PDS) and we do 
know that solar flares results from reconnection of a magnetic field
with a complex structure.  Lu and Hamilton (1991) described power law
distributions of flare energy release with a cellular automata model which
is also a kind of a near-critical pulse avalanche.
      
 Summarizing the issue:
 
 The time variability of GRBs can hardly  appear straightforwardly as a 
result of internal shock collisions or as a consequence of variations of the
external 
medium. The time behavior should be associated with a dynamical process   
that makes the outflow strongly inhomogeneous in a wide range of scales
giving rise to a kind of fractal pattern.

  The inhomogeneous structure of the outflow removes the main objections   
against a single explosion scenario. An argument in favor of the recurrent central 
engine is the absence of evident evolution of long events (Fenimore 1999). 
However,
some evolution probably exists (e.g., Ford et al. 1995) and this 
argument can hardly be used as a proof.

\section{Lorentz Factor, Compactness and Emission Regime}

 The cosmological origin of GRBs unavoidably implies a relativistic motion of 
the emitting region towards the observer.
 Let us consider an emission episode with a luminosity, $L = 10^{50}$ erg/s,  
and a characteristic variability time scale, 1~s. The size of the emitting 
region 
should not be greater than 1 light second, i.e., $r \sim 3 \cdot 10^9$cm. 
Then, assuming no relativistic motion we obtain:

 A compactness parameter:  $$\ell = \frac{L \sigma_T}{m_e c^3 r} \sim 10^{12}$$

 An equilibrium (blackbody) temperature $T = (L/4\pi r^{2} \sigma)^{1/4} 
\sim 30$ keV.

 We can hardly  see anything except the 30 keV Planck spectrum using this  
assumption
and such a system cannot be stable - it should explode.  Now let us
describe the emission region as a blob, quasi-spherical in the comoving frame,
moving towards the observer with a Lorentz factor $\Gamma$. Then the comoving
luminosity is $L_c = L \Gamma^{-4}$ ($\Gamma^{-2}$ from angular collimation,
$\Gamma^{-1}$ from time transformation and $\Gamma^{-1}$ from blueshift), 
where $L$ is the apparent  luminosity (assuming isotropy).

 For the size of the emission region we take $r_c = r \Gamma$.
Then the comoving compactness is:

\begin{equation}
 \ell_c = \ell \Gamma^{-5} = 10^{12} \cdot 
\Gamma^{-5}
\end{equation}

 If we want to deal with simple linear physics describing the gamma ray
emission, we should take $\Gamma > 100$. Then we have no problem with 
intense pair production and can apply the optically thin synchrotron-self 
Compton models (see,  e.g., Panaitescu \& M\'esz\'aros 1998). 
This is the most popular approach and the constraint $\Gamma >
100$ is generally accepted.  

 For other comoving values we have:

 The energy density at the emitting surface:

\begin{equation}
\epsilon_c \sim 3 \cdot 10^{19}  \Gamma^{-6}\quad {\rm erg/cm}^3.
\end{equation}

The equipartition magnetic field:

\begin{equation}
H_c \sim 3 \cdot 10^{10}  \Gamma^{-3}\quad {\rm G}
\end{equation}

The equilibrium temperature:

\begin{equation}
T_c \sim 30 \cdot  \Gamma^{-3/2}\quad {\rm keV}. 
\end{equation}

And the temperature, blueshifted to the observer frame:

\begin{equation}
T \sim 30 \cdot  \Gamma^{-1/2}\quad {\rm keV}. 
\end{equation}

 The global size of the relativistic fireball (or the distance
from the source, having in mind a jet geometry), for a characteristic emission 
time of 300 s:

\begin{equation}
R \sim 10^{13} \Gamma^2\quad {\rm cm}.
\end{equation}

 One can obtain a large variety of physical conditions depending on
the Lorentz factor. On the other hand, there are arguments for a small
dispersion of the Lorentz factor in different GRBs (e.g., a sharp break in the
average PDS, Beloborodov, Stern \& Svensson 1999). What is the typical 
Lorentz factor? This is one of the most important issues in the whole
GRB problem.

 At a huge Lorentz factor, $\Gamma \sim 1000$, the blast wave passes a distance 
 of order of a parsec  during the emission phase. 
 This value was assumed in the model of Shaviv \& Dar (1995b)
describing GRB emission as upscattering of the star light by a 
$\Gamma \sim 1000$ blast wave crossing a globular star cluster. 
The model gives a wrong description of the GRB time variability 
(e.g., fact 2.3 can not be explained). It seems that we do not need such  
a Lorentz factor for any other 
purposes and taking into account some other problems (e.g., the requirement of a   
good vacuum, $n < 10^{-5} {\rm cm}^{-3}$ for long events), 
we will not consider this possibility seriously.

 The main choice is between large ($\Gamma \sim 100 - 300$)
and moderate ($\Gamma \sim 10 - 50$) Lorentz factors. This choice will define 
the emission regime: in the first case this should be optically thin 
synchrotron (synchrotron - self Compton), in the second case, intensive 
pair production should take place and we have a much more complicated 
nonlinear, optically thick emitting system.

\section{$\Gamma \sim 100 - 300$ versus $\Gamma \sim 10 - 50$ or Optically 
Thin versus Optically Thick Emission}

  A large Lorentz factor ($\Gamma \sim 300$) is attractive because it 
can explain the GRB emission as a result of the interaction of the blast wave
with the interstellar medium. Indeed, the kinetic energy of the 
interstellar
gas swept up by the fireball with a Lorentz factor $\Gamma$ at the observer 
time $t$ is

\begin{equation}
E_{KE} = 6 \cdot 10^{49} t_{100}^3 \Gamma_{100}^7 \cdot n  \quad {\rm  erg},
\end{equation}
\noindent
 where $n$ is the gas density in cm$^{-3}$, $t_{100} = t/100$ s, and
 $\Gamma_{100} =
\Gamma /100$. Accepting the value $t=100$ s for the emission phase 
(for a recurrent central engine
model one can afford a slightly smaller $t$; for a single explosion model 
one must take $t > 100$ s in some cases) and $n = 0.1$ cm$^{-3}$ 
we obtain $E_{KE} \sim 10^{52}$ erg for $\Gamma = 300$. Under such conditions 
we should see a strong energy dissipation from the interaction between 
the fireball and the 
interstellar medium within the first 100 s. If $n= 10^{-4}$ cm$^{-3}$ 
(a GRB in a galactic halo) then one can slightly adjust $T$ and $\Gamma$ to 
obtain a considerable deceleration of the fireball in a reasonable time.  

 The interaction between the fireball and the external medium solves 
the free energy
problem. The free energy source for the gamma ray emission is just the bulk
kinetic energy of the fireball. The most popular scheme of the emission 
is shock particle acceleration and synchrotron - self Compton radiation 
(e.g., Tavani 1996; Panaitescu \& M\'esz\'aros 1998;  Dermer 1998)
One can see from Eq.~(2) that pair production is negligible at such high 
$\Gamma$ and the electron scattering optical depth is small. Therefore we 
deal with 
optically thin linear emission. The involved 
physics is well studied and easy to work with. However we have a number of 
very serious problems with this simple linear physics. 

 The first difficult question is {\bf ``what causes the specific time variability 
of GRBs?''}. Is it inhomogeneities of the external medium? How can one then
explain the stretched exponential average time profile and the power law PDS?  
With a fractal structure of 
the interstellar medium? Hhow can we then explain the huge amplitude of
variations (Fig.~2)? We certainly need some essentially nonlinear system
to produce rapid variations by 3 orders of magnitude.

 A process which can produce both a large dynamical range of variations and
a wide range of time scales is magnetic reconnection (we note that it works
in a similar way in solar flares). An equipartition magnetic field with a 
complex geometry can be generated by a turbulent dynamo. Then such a field 
can gain additional energy with compression in the deceleration stage and 
reconnect. This could be a solution of the problem of time variability 
for $\Gamma \sim 100 - 300$
(we are unfortunately not able to solve this problem at a quantitative level).

The next problem arises from the {\bf GRB spectra.}
 In both variants of energy release at $\Gamma \sim 100 - 300$, shock 
acceleration and magnetic 
reconnection, the gamma-ray emission is blueshifted optically thin 
synchrotron radiation. 

Real GRB spectra are well approximated by the Band 
expression (Band et al. 1993) consisting of two asymptotic power laws: 

$$dN/dE \propto E^\alpha$$ 

at small $E$, $$dN/dE \propto E^\beta$$ at large $E$
and $$dN/dE \propto E^\alpha e^{-E/E_p}$$ in the intermediate range. $E_p$ 
parameterizes the break energy.
At large negative $\beta$,  this expression resembles the hard X-ray 
spectra of AGNs, especially if one 
subtracts the reflection hump (see Zdziarski et al. 1997). $E_p$ is 
associated with the
pair temperature in that case. In AGNs, we have $\alpha$ close to $-2$
($-1.9$ is the most typical value, and $E_p \sim 60 - 150$ keV). The high energy 
spectrum, $E \gg E_p$, in AGNs cannot be reconstructed because of 
poor photon statistics.
  
In GRBs 
the soft part is considerably harder: $\alpha$ varies between  
$-2$ and $+1$ (Band et al. 1993).  There are some fits of spectra with 
$\alpha \sim +1$ but they 
have large errors, a short fitting interval, and a low $E_p$ (Crider et al. 
1997). 
The largest $\alpha$
that  one can trust is near 0  (Preece 1998, private communication). 
$E_p$ is also 
larger than that for AGNs and variable within a single burst. 
The highest values of $E_p$
is above the BATSE range ($\sim 1.5$ MeV), the lowest is below the BATSE range 
($\sim 30$ keV) and for the main fraction of spectra, 100 keV $< E_p <$ 500 keV
(Band et al. 1993). The typical hard energy slope is $-2.8  < \beta < -1.7$
(clustering around $\beta \sim -2.1$, Preece et al. 1996),
sometimes much steeper, consistent with a pure exponential cutoff 
($\beta \sim -\infty$).  

Summarizing the GRB spectral phenomenology:

- The GRB low energy (hard X-ray) spectra are considerably harder then the 
AGN spectra and have a break at a higher energy.  

- The GRB spectra are much more diverse than the AGN spectra, nevertheless 
they have a typical shape: a harder low energy power law, an exponential break,
and a softer high energy power law. 

- The spectra evolve during a single pulse. A pulse starts with a maximum 
$E_p$, then $E_p$ decreases, sometimes by factor of a few (Ford et al. 1995). 

 How do optically thin synchrotron models fit this spectral pattern?
The first problem appears with the low energy spectra. A synchrotron 
model cannot give a spectrum with $\alpha > -2/3$, while there are considerably
harder spectra, $\alpha = 0$, at least. This issue is studied by 
Preece et al. (1998).
The second problem is the spectral break, sharp enough to be fitted with  
an exponential (Band et al. 1993). It implies a very sharp electron energy 
distribution, 
which remains sharp during rapid evolution (note that the synchrotron 
photon energy is proportional to the square of the electron Lorentz factor). 

 From my point of view these problems are fatal for the synchrotron shock 
models. We should search for a less linear and less trivial physics 
to explain GRB emission (especially if we want to explain the nontrivial
time variability at the same time).

 I started the discussion with a comparison between GRB and AGN spectra 
and this is more motivated than it could seem at first sight. There is
a number of arguments
that in the case of GRBs as well as in the case of AGNs that we deal with an
equilibrium $e^+e^-$ pair plasma. It is surprising that while there exist 
 a large number 
of works on synchrotron shock models, we know of very few attempts to describe
GRB spectra with a Comptonizing pair plasma. I can  only mention the works of  
Ch.~Thompson (see Thompson 1998 and references therein) 
and Ghisellini \& Celotti (1999). Liang (1997) and Liang et al. (1997) studied 
optically thick thermal Comptonization in application to GRBs taking
the temperature and the optical depth as external parameters.

   It is a well known fact that the pair plasma is a good thermostat and
is able to produce spectra with a stable break (which also can be sharp)  
in the X-ray range (Svensson 1984). The break results from quasi-thermal 
Comptonization.  Its position is defined by the pair equilibrium
and depends on the compactness parameter and on the type of the energy supply:
pure thermal (direct heating of Maxwellian electrons), 
nonthermal (heating of the relativistic 
tail  of electron energy distribution), or hybrid (both).

In the pure thermal case, the pair temperature is self-adjusted in a way to 
support pair production at the tail of the photon energy distribution.
The resulting temperature decreases logarithmically with increasing 
compactness, at $\ell \sim 1000$ the pair temperature is $\sim 40$ keV
and the peak in $\nu F_{\nu}$ distribution appears at $\sim 80$ keV. 
This is the energy in the comoving frame, and it implies a too 
small Lorentz factor as the average observable $E_p$ is 300 -- 400 keV.

 A smaller pair temperature can be achieved in nonthermal or hybrid model.
To demonstrate that it is in principle possible to reproduce GRB spectra with optically 
thick pair plasma, I made a series of simulations with a large particle 
nonlinear Monte-Carlo code (Stern et al. 1995) for high compactnesses
($\ell \sim 1000 - 2000$). Figure 5 demonstrates the result of one attempt that
can be considered as more or less successful. The break position at 20 -- 30
keV is consistent with a Lorentz factor 10 -- 20 which implies a higher 
compactness (see eq. 2) which in turn would give cooler pairs. The simulation at 
$\ell \gg 1000$ is technically difficult.  The impression is that  
consistency with data can be achieved at $\ell = 10^4$ and $\Gamma \sim 20 - 30$.

\begin{figure}[t]
\centerline{\epsfig{file=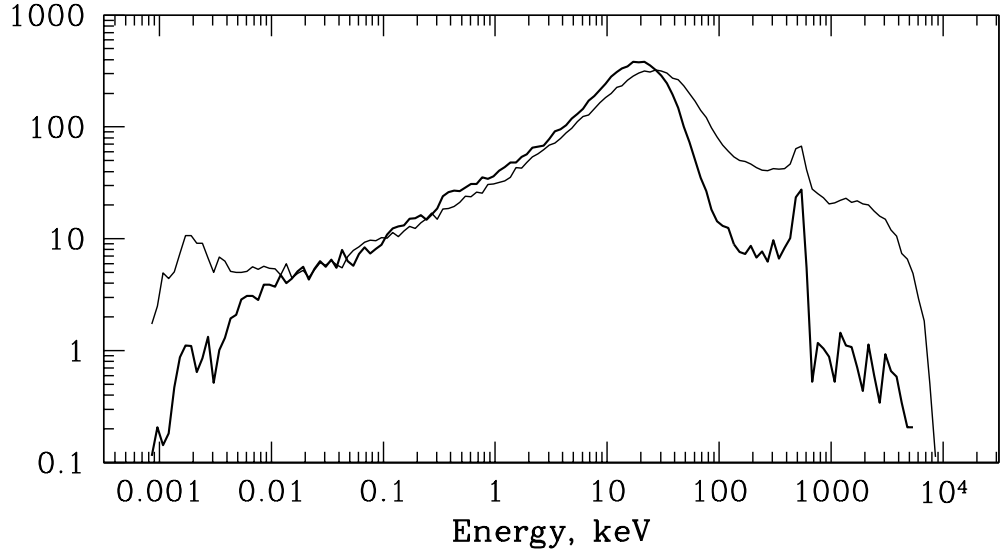,width=14cm}}
\caption{Examples of simulated spectra of optically thick pair plasma. The 
nonlinear large particle Monte-Carlo technique of Stern et al. (1995) was 
used ($2^{17}$ large particles). The active 
region is a sphere with uniform energy injection, while the 
temperature and the pair density depend on radius (10 discrete shells): both 
are obtained in the simulation as a result of the pair and energy 
balance. Energy is 
injected by instant acceleration of pairs to 10 MeV. The synchrotron emission is 
strongly self-absorbed (the peak near 0.001 keV is the harder edge of the 
partially self-absorbed synchrotron spectrum), the energy of the accelerated 
pairs goes 
to photon Comptonization, or heats thermal electrons through synchrotron  
emission - self-absorption.
 The compactness is 1000, the magnetic field $10^6$G. The pair optical depth is
$\sim 10$, and the temperature varies from 4 keV in the center to 
10 keV near the surface. The two spectra correspond to different 
states of the system. The thin line shows the steady-state spectrum. 
The thick line shows
the decaying state. The spectrum is integrated over time 
from $2 r/c$ to $4 r/c$ after the energy injection 
was turned off. The pair depth dropped in the second case and we see
radiation escaping from the center where the temperature is lower.
}
 \label{fig-5}
\end{figure}

The recipe how to obtain a proper spectral shape can be formulated as follows:

The main condition to obtain a hard spectrum below $E_p$ is photon 
starvation, i.e., only a small number of soft photons enters the Compton  
upscattering process (see Zdziarski, Coppi, \& Lamb 1990). The main source of soft photons is 
synchrotron radiation of the nonthermal pair component. To get rid of it
one should restrict the nonthermal tail to the energy range for which the 
synchrotron radiation is  reabsorbed by thermal pairs. In the 
presented example, this condition is fulfilled and the resulting low 
energy spectral slopes, $\alpha = -1.1$ for the steady-state spectrum 
and $\alpha = -0.95$ for the decaying state, are typical for GRBs. To obtain a 
harder spectrum one should take a larger magnetic field and energy density 
to have a higher reabsorption energy. A rising pair optical depth and energy 
density will eventually lead to a Planck spectrum with the temperature estimated
by Eq.~(6).

An optically thick pair plasma provides another advantage: a nonlinearity     
which can give rapid variations with a large amplitude. A large pair 
optical depth can be generated during a few $r/c$ and can annihilate on 
the same time scales, i.e., when the emitting system turns off, it does not just 
cool down -- it disappears.

 The only objection against a moderate Lorentz factor and a large compactness is 
associated with the GeV photon emission detected in some bursts. At a high
compactness, high energy photons should be absorbed through photon-photon 
pair production. This constraint can be easily avoided by  assuming that
the hundred keV -- MeV  emission and the GeV emission originate from different  
processes in different places, e.g., the latter could result from shock 
acceleration, the former from magnetic reconnections behind the shock.

%\newpage

Summarizing the issue:

 A large Lorentz factor ($\Gamma \sim 100 - 300$) naturally enables the 
conversion of
the fireball kinetic energy into radiation through interaction with external 
matter. It implies a simple, linear mechanism of gamma ray emission which
does not seem to satisfy the data.

 At a moderate Lorentz factor  ($\Gamma \sim 10 - 30$),  more 
interesting physics appear: a nonlinear system of an optically thick pair 
plasma and radiation at a high compactness. This regime is much more 
difficult to study. Nevertheless, this case can hopefully provide a welth 
on nonlinear phenomena that could explain many puzzling properties of GRBs.

\acknowledgments

 Author is grateful to Juri Poutanen and Roland Svensson for useful 
 discussions. I thank Jana Tikhomirova for assistance.
This work is supported by the Wennergren Foundation for Scientific Research, a
Nordita Nordic Project, and the Swedish Royal Academy of the Sciences.

\end{document}